\documentclass[12pt]{article}
\def \bea{\begin{eqnarray}}
\def \beq{\begin{equation}}

\def \bt{\bar t}

\def \eea{\end{eqnarray}}
\def \eeq{\end{equation}}

\def \hp{\hat{p}}
\def \ket#1{| #1 \rangle}

\def \mat#1#2{\langle #1 | #2 \rangle}

\def \od{\overline{D}^0}

\def \pr{\parallel}
\def \s{\sqrt{2}}

\begin{document}

\begin{flushright}
TECHNION-PH-2001-19\\
EFI 01-07 \\
hep-ph/0103110 \\
March 2001 \\
\end{flushright}

\medskip
\begin{center}
\large
{\bf Measuring $D^0$--$\od$ Mixing and Relative Strong Phases}
\end{center}
\begin{center}
\large
{\bf at a Charm Factory}
\footnote{To be submitted to Physics Letters B.}
\end{center}

\bigskip

\begin{center}
\normalsize
{\it Michael Gronau} and {\it Yuval Grossman} \\

\medskip
{\it Department of Physics, Technion-Israel Institute of Technology \\
Technion City, 32000 Haifa, Israel}

\bigskip
and
\bigskip

{\it Jonathan L. Rosner \\
\medskip

Enrico Fermi Institute and Department of Physics \\
University of Chicago, Chicago, Illinois 60637 } \\

\bigskip
{\bf ABSTRACT}

\end{center}

\begin{quote}
We propose a set of measurements at a charm factory to determine the
$D^0$--$\od$ mixing parameters $\Delta M$ and $\Delta\Gamma$ and the
strong phase difference $\delta$ between doubly-Cabibbo-suppressed (DCS)
and Cabibbo-favored (CF) neutral $D$ decays into $K^-\pi^+$.  The method
can also be used to measure strong phase differences between other
corresponding DCS and CF amplitudes.  These phase differences are
important for studies of $D^0$--$\od$ mixing.
\end{quote}
\bigskip

The time development of decays of neutral $D$ mesons to
doubly-Cabibbo-suppressed (DCS) modes such as $K^+\pi^-$ exhibits
interesting behavior.  At $t=0$ the only term in the amplitude is
the direct DCS $D^0 \to K^+ \pi^-$ term, but for $t>0$ a
$D^0$--$\od$ mixing contribution appears.  The interference
of this term with the DCS contribution involves the lifetime
and mass differences of the neutral $D$ mass eigenstates as well
as the final-state strong phase difference $\delta$ between
the Cabibbo-favored (CF) $\od \to K^+ \pi^-$ and DCS $D^0 \to K^+ \pi^-$
decay amplitudes.

A recent study of this process by the CLEO Collaboration \cite{CLDCS},
when combined with direct measurements of the lifetime difference
between CP-even and CP-odd neutral $D$ mesons \cite{E791,FOCUS,BElife,CLlife,%
BeBCP},
suggests that this strong phase difference may be large \cite{Berg}.  The
FOCUS Collaboration has also presented data in agreement with CLEO's
results \cite{FOCD}.
Several attempts to estimate this phase theoretically \cite{CC,BP,FNP,Uspin}
involve various degrees of model-dependence. A recent proposal to
measure the phase \cite{GolP} assumes that $\Delta I =0$ and $\Delta I =1$
DCS $D \to K\pi$ amplitudes into $I=1/2$ final states involve equal strong
phases. In the present note we suggest an entirely assumption-free method
for determining this phase experimentally.  We show that related measurements
are capable of determining $D^0$--$\od$ mixing parameters at the percent
level.

A charm factory operating at the $\psi(3770)$ resonance produces $D^0 \od$
pairs in a state of definite charge-conjugation eigenvalue $C = -$. At slightly
higher energies one can produce such pairs with $C = +$ \cite{GolR}.
One may tag one of the neutral $D$ mesons as a CP eigenstate through its decays
into CP eigenmodes, such as $K_S(\pi^0,\rho^0,\omega,\eta,\eta',\phi)$,
$K^+ K^-$, and $\pi^+ \pi^-$.
(These decays are important in a different context of studying the weak
phase $\gamma$ in $B\to D^0 K$ \cite{GW,CLEOGW}.)
The other neutral $D$ meson must then have opposite CP if $C(D^0 \od) = -$ and
the same CP if $C(D^0 \od) = +$ \cite{GolR}.
One measures its
decay rate into $K^-\pi^+$, which includes an interference between CF and
DCS amplitudes.  The measured rate thus is given in terms of the
already-measured CF and DCS rates and the relative strong phase $\delta$,
which permits a determination of $\delta$.
(This phase plays an important role in a method \cite{ADS} for
measuring $\gamma$ in  $B\to D^0 K$.)

We now discuss details of this method, beginning with conventions and
notation and ending with an estimate of the achievable precision.
For the majority of our discussion we shall neglect CP violation in neutral
$D$ mixing and decays, which is expected to be very small in the Standard
Model.
Our master equations for correlated hadronic decays of $D^0 \od$ pairs,
Eqs.~(\ref{f1f2-})--(\ref{f1f2+}) below, apply also to the case of CP
violation.
Towards the end of our study we will argue that including CP violation beyond
the Standard Model has a negligible effect on the proposed measurement of the
relative strong phase $\delta$.

The mass eigenstates in the neutral $D$ meson system may be defined as
\bea
D_1 & \equiv & p \ket{D^0} + q \ket{\od}~~, \nonumber \\
D_2 & \equiv & p \ket{D^0} - q \ket{\od}~~,
\eea
where $|p|^2+|q|^2=1$,
with corresponding eigenvalues $\mu_{1,2} \equiv M_{1,2} - i \Gamma_{1,2}/2$.
Neglecting CP violation in $D^0$--$\od$ mixing, we adopt the convention
\cite{Berg} in which $D_1$ is the CP-odd and $D_2$ the CP-even state, choosing
further $p = q = 1/\s$. We define
\beq
M \equiv \frac{M_1 +M_2}{2}~~,~~~\Gamma \equiv \frac{\Gamma_1 + \Gamma_2}{2}~~,
~~~\mu \equiv \frac{\mu_1 +\mu_2}{2}~~~, \nonumber
\eeq
\beq
x \equiv \frac{M_2 - M_1}{\Gamma}~~,~~~y \equiv \frac{\Gamma_2 - \Gamma_1}
{2 \Gamma}~~,~~~\Delta \mu \equiv \mu_2 - \mu_1 = \Gamma(x-iy)~~.
\eeq
The decay amplitudes are defined as
\beq \label{eqn:amps}
\mat{K^- \pi^+}{D^0} \equiv A e^{i \delta_R}~~,~~~
\mat{K^- \pi^+}{\od} \equiv \bar A e^{i \delta_W}~~,~~~
\delta \equiv \delta_R - \delta_W~~,
\eeq
where $\delta_R$ and $\delta_W$ refer, respectively, to ``right-sign''
and ``wrong-sign'' strong phases in the $K^- \pi^+$ system.

With the above-mentioned convention for CP eigenstates, self-consistency
requires us to take $D^0 \equiv c \bar u$, $\od \equiv - u \bar c$ as in
Ref.~\cite{GHLR}.  The decay $D_1 \to K^+K^-$ then is forbidden.  With
this convention, in the SU(3) limit, we would have $\bar A/A = - V_{cd}
V^*_{us}/ V_{ud}V^*_{cs} = +\tan^2 \theta_C \simeq 0.05$, where $\theta_C$
is the Cabibbo angle.  We thus take, more generally, $\bar A/A \equiv r > 0$.

We note that (\ref{eqn:amps}) implies
$\mat{K^+ \pi^-}{D^0} = - \bar A^*e^{i \delta_W}$ which is equal, by a
U-spin $s \leftrightarrow d$ substitution and a replacement of CKM
elements, to $[V_{cd}V^*_{us}/ V_{ud}V^*_{cs}]A^* e^{i \delta_R}$.
Thus $\delta = 0$ in the SU(3) limit \cite{KTWZ,Wolf}.

We begin by considering the process studied by CLEO and FOCUS
\cite{CLlife,FOCD}, which does not require a symmetric charm factory.
A state which is identified at the time of production as a $D^0$, e.g.,
by the decay $D^{*+} \to \pi^+ D^0$, evolves in time as $D^0(t) = D^0 f_+(t)
+ \od f_-(t)$, where
\beq
f_\pm(t)= \frac{1}{2}[e^{-i \mu_1 t}\pm e^{-i \mu_2 t}] = e^{-i \mu t}
\left\{ \begin{array}{c} \cos (\Delta \mu t/2) \\
i \sin(\Delta \mu t/2) \end{array}\right\}~~.
\eeq
Similarly $\od (t) = \od f_+(t)+D^0 f_-(t)$.  The time-dependence of
a ``wrong-sign'' final state is governed by the amplitude
$$
\mat{K^- \pi^+}{\od (t)} = e^{-i \mu t} [ \mat{K^- \pi^+}{\od}
\cos(\Delta \mu t/2) + \mat{K^- \pi^+}{D^0}i \sin(\Delta \mu t/2)]
$$
\beq
\simeq \mat{K^- \pi^+}{D^0} e^{-i \mu t} e^{i \delta_W}
[r + i e^{i \delta} \Gamma t (x - iy)/2]~~,
\eeq
where the small-argument approximation is sufficient for the trigonometric
functions since $r,x,y \ll 1$.  The ratio of wrong-sign to right-sign decays,
as a function of time, is then
$$
R(t)\equiv \left|\frac{\mat{K^- \pi^+}{\od (t)}}{\mat{K^- \pi^+}{D^0(t)}}
\right|^2 \simeq \left| r +i \frac{\Gamma t}{2}(x-iy) e^{i \delta} \right|^2~~.
$$
\beq \label{eqn:R}
= r^2 +r y' \Gamma t +\left( \frac{\Gamma t}{2}\right)^2 (x^2 +y^2)~~,
\eeq
where $x' \equiv x \cos \delta +y \sin \delta$, $y' = y \cos \delta -
x \sin \delta$.  The suggestion of large $\delta$ first arose because
the CLEO analysis \cite{CLDCS} slightly favored negative $y'$ in the term
linear in $t$ in the above equation, while the FOCUS result for $y$
\cite{FOCUS}, the dominant one in terms of statistical error, favored
$y>0$. The results for $y$ are summarized in Table 1.  The evidence for
$y > 0$ now is slightly less than $2 \sigma$.  CLEO reported $y' =
(-2.5^{+1.4}_{-1.6} \pm 0.3)\%$ in an analysis allowing for CP violation
and $y' = (-2.3^{+1.3}_{-1.4} \pm 0.3)\%$ when CP symmetry was assumed.

\begin{table}
\caption{Results on $y$. A shorter lifetime for the CP-even neutral
$D$ corresponds to $y >0$.}
\begin{center}
\begin{tabular}{c c c} \hline
Reference & Value (percent) & $\Delta \chi^2$ \\ \hline
E791 \cite{E791}    &  $0.8 \pm 2.9 \pm 1.0$       & 0.16 \\
FOCUS \cite{FOCUS}  & $3.42 \pm 1.39 \pm 0.74$     & 0.80 \\
Belle \cite{BElife} &  $1.0^{+3.8+1.1}_{-3.5-2.1}$ & 0.06 \\
CLEO \cite{CLlife}  & $-1.1 \pm 2.5 \pm 1.4$       & 1.18 \\
Belle \cite{BeBCP}  & $1.16^{+1.67}_{-1.65}$       & (a)  \\ \hline
Average             &      $2.0 \pm 1.2$           & 2.2  \\ \hline
\end{tabular}
\end{center}
\centerline{(a) Not used in average since no systematic error was quoted.}
\end{table}

If all three terms in Eq.~(\ref{eqn:R}) (constant, linear, and quadratic in
$t$) are measurable, and $y$ is also measured, then it is possible in
principle to determine $x$ and $\delta$ up to discrete ambiguities.  However,
most (but not all)
estimates of $x$ and $y$ within the Standard Model are considerably less than
a percent \cite{Nelson}.  If this is so, the last term in (\ref{eqn:R}) may
be inaccessible even though there may exist evidence for the $r y'$ term,
and we need an independent determination of $\delta$.  This may be achieved
through experiments at a charm factory, as we now show.

We consider a $D^0 \od$ pair produced in a state of
definite charge-conjugation eigenvalue $\eta_C = \pm 1$.  The initial
states $\Psi_{\eta_C}$, namely,
\beq
\Psi_\pm = \frac{1}{\s} [D^0(\hp) \od(-\hp)\pm \od (\hp)D^0(-\hp)]~~,
\eeq
evolve in time to
$$
\Psi_\pm(t,\bt) = \frac{1}{\s} e^{-i \mu (t+\bt)} \left\{ \cos[\Delta \mu
(t \pm \bt)/2] [D^0(\hp) \od(-\hp)\pm \od (\hp)D^0(-\hp)] \right.
$$
\beq
\left. \pm i \sin[\Delta \mu (t \pm \bt)/2] [D^0(\hp)D^0(-\hp) \pm
\od (\hp) \od (-\hp)] \right\}~~,
\eeq
Here $t$ refers to the proper time of a state traveling along the $+\hp$
direction, while $\bt$ refers to one traveling along $-\hp$.

We now consider decays of these correlated systems into various final states,
searching in particular for interference effects depending on $\delta$.
In all cases we integrate with respect to proper time, since vertex
separation in a symmetric $e^+e^-$ ``charm factory'' is likely to be
problematic.
An early study of correlated $D^0 \od$ decays into specific flavor final
states, assuming $\delta =0$, was carried out by Bigi and Sanda \cite{BS}.
More recently Xing \cite{Xing} has considered both time-dependent and
time-integrated decays into correlated pairs of states, including some effects
of a nonzero final state phase difference. However, he has not considered the
cases ($3^-$) and ($3^+$) below, from which we propose to measure $\delta$.
For completeness, we derive general expressions for time-integrated decay
rates
into a pair of final states $f_1$ and $f_2$, from $C = -1$ and $C = +1$ $D^0
\od$ states, in agreement with \cite{Xing}:
\bea\label{f1f2-}
\Gamma^{C=-1}(f_1,f_2) & = & \frac{1}{2}|A^{(-)}|^2 \left[
\frac{1}{1-y^2} + \frac{1}{1+x^2} \right] \nonumber \\
& + & \frac{1}{2} |B^{(-)}|^2 \left[
\frac{1}{1-y^2} - \frac{1}{1+x^2} \right]~~,\\
\label{f1f2+}
\Gamma^{C=+1}(f_1,f_2) & = & \frac{1}{2}|A^{(+)}|^2 \left[
\frac{1+y^2}{(1-y^2)^2} + \frac{1-x^2}{(1+x^2)^2} \right] \nonumber \\
& + & \frac{1}{2} |B^{(+)}|^2 \left[
\frac{1+y^2}{(1-y^2)^2} - \frac{1-x^2}{(1+x^2)^2} \right] \nonumber \\
& + & 2{\rm Re} \left\{ A^{(+)*}B^{(+)} \left[ \frac{y}{(1-y^2)^2}
+ \frac{ix}{(1+x^2)^2} \right] \right\}~~,
\eea
where
\bea\label{A+-}
A^{(\pm)} &=& \mat{f_1}{D^0}\mat{f_2}{\od} \pm \mat{f_1}{\od}\mat{f_2}{D^0}~~,
\\
\label{B+-}
B^{(\pm)} &=& \mat{f_1}{D^0}\mat{f_2}{D^0} \pm \mat{f_1}{\od}\mat{f_2}{\od}~~.
\eea

The rate expressions simplify if one of the states (say, $f_2$) is a CP
eigenstate $S_\zeta$ with eigenvalue $\zeta = \pm 1$:
\beq
\Gamma^{C=-1}(f_1,S_\zeta) = |A_{S_\zeta}|^2 \frac{|\mat{f_1}{D^0} +\zeta
\mat{f_1}{\od}|^2}{1-y^2}~~,
\eeq
\beq
\Gamma^{C=+1}(f_1,S_\zeta) = |A_{S_\zeta}|^2
\frac{|\mat{f_1}{D^0} -\zeta \mat{f_1}{\od}|^2}{(1 + \zeta y)^2}~~,
\eeq
where $A_{S_\zeta}=\mat{S_\zeta}{\od}$,
and we have used $CP \ket{D^0} = - \ket{\od}$.

The expressions for products of amplitudes in Eqs.~(\ref{A+-}) and (\ref{B+-})
can be easily generalized allowing for CP violation:
\bea
A^{(\pm)} &=& {p \over q} A_1 A_2
\left(\lambda_2 \pm \lambda_1\right) ~~,\\
B^{(\pm)} &=& {p \over q} A_1 A_2
\left(1 \pm \lambda_2 \lambda_1\right) ~~,
\eea
where
\beq\label{lambda}
A_i \equiv \mat{f_i}{D^0}~~,~~~
\bar A_i \equiv \mat{f_i}{\od}~~,~~~
\lambda_i \equiv {q \over p}{\bar A_i \over A_i}~~.
\eeq
\medskip
Here and everywhere we use a normalization in which $\Gamma(D^0 \to f_i) =
|A_i|^2$.

Keeping terms up to order $r^2, x^2, y^2$ in the expressions for rates,
and assuming CP conservation
we list the following results for various cases:

\medskip
\leftline{\underline{\it $C = -1$ $D^0 \od$ states:}}
\medskip

($1^-$) \underline{$K^- \pi^+(\hp) K^- \pi^+(-\hp)$.}
\bea
\Gamma^{(-)}(K^-\pi^+, K^-\pi^+) & = &
\frac{1}{2}A^4|1 - r^2 e^{-2i\delta}|^2(x^2 + y^2)
\nonumber\\
& \approx & \frac{1}{2}A^4(x^2 + y^2)~~,
\eea
where $A$ was defined in Eq.~(4).
This process serves to measure mixing effects when normalized by the one which
follows.
\medskip

($2^-$) \underline{$K^- \pi^+(\hp) K^+ \pi^-(-\hp)$.}
\bea
\Gamma^{(-)}(K^-\pi^+, K^+\pi^-) & = &
A^4|1-r^2 e^{-2i\delta}|^2[1-\frac{1}{2}(x^2-y^2)]
\nonumber \\
& \approx & A^4[1-2r^2\cos 2\delta - \frac{1}{2}(x^2 - y^2)]~~.
\label{eqn:rr}
\eea
The dominant contribution is proportional to $A^4$, so by comparing this case
with the previous one we can learn the combination $x^2 +y^2$ describing
mixing.  Similar information could be gleaned from a fit to the
time-distribution of a single tagged $D^0$ as described above, if $x^2 +y^2$ is
sufficiently large, but this may not be the case. The small interference terms
in (\ref{eqn:rr}) are probably unmeasurable.
\medskip

($3^-$) \underline{$K^- \pi^+(\hp) S_{\zeta}(-\hp)$.}
\bea \label{3-}
\Gamma^{(-)}(K^-\pi^+, S_\zeta) & = &
A^2 A_{S_\zeta}^2 |1 + \zeta r e^{-i\delta}|^2(1 + y^2)
\nonumber\\
& \approx & A^2 A_{S_\zeta}^2(1 + 2 \zeta r \cos \delta)~~.
\eea
Recall $r \approx \tan^2 \theta_C \simeq 0.05$ in the SU(3) limit.  By
comparing rates with $\zeta = +1$ final states such as $K^+K^-$ and
$\zeta = -1$ final states such as $K_S(\rho^0,\omega,\phi)$
one can measure the ratio $(1 + 2 r \cos \delta)/(1 - 2 r \cos \delta)$
and, given an independent measurement of $r$, one can obtain $\cos \delta$.
\medskip

($4^-$) \underline{$K^- \pi^+(\hp)  \ell^- (-\hp)$.}

Using a leptonic $\od$ flavor tag and defining
$A_{\ell^-}=\mat{\ell^- X}{\od}$, one finds
\beq
\Gamma^{(-)}(K^-\pi^+, \ell^- )  =
A^2 A_{\ell^-}^2 [1 - \frac{1}{2}(x^2 - y^2)]~~.
\eeq
This process serves as a normalization for the one which follows
describing the opposite-sign $D^0$ flavor tag.
\medskip

($5^-$) \underline{$K^- \pi^+(\hp)  \ell^+ (-\hp)$.}
\beq
\Gamma^{(-)}(K^-\pi^+, \ell^+ )  =
A^2 A_{\ell^+}^2 [r^2 + \frac{1}{2}(x^2 + y^2)]~~,
\eeq
where $A_{\ell^+}=\mat{\ell^+ X}{D^0}=A_{\ell^-}$.
By comparing this process with the previous one, we obtain $r^2 + (x^2 + y^2)
/2$.
This may be of interest for mixing parameters if $r,x,y$ are of comparable
size but, as mentioned, it is much more likely that $x,y \ll r$ in which case
this process can be used to measure $r$.
\medskip

($6^-$) \underline{$ S_\zeta(\hp)  \ell^+ (-\hp)$.}
\beq
\Gamma^{(-)}(S_\zeta, \ell^+ ) =  A_{\ell^+}^2 A_{S_\zeta}^2
(1 + y^2)~~.
\eeq
Strictly speaking, the $y^2$ correction is one order higher
in small parameters
than we have been keeping, since $A_{S_\zeta}^2$ is
already of order $r$.  This process serves as a normalization for others.
\medskip

\leftline{\underline{\it $C = +1$ $D^0 \od$ states:}}
\medskip

($1^+$) \underline{$K^- \pi^+(\hp) K^- \pi^+(-\hp)$.}
\beq\label{1+}
\Gamma^{(+)}(K^-\pi^+, K^-\pi^+) =
4A^4[r^2 + ry' + \frac{3}{8}(x^2 + y^2)]~~.
\eeq
This expression gives information similar to that learned from the
time-dependence (\ref{eqn:R}) in the decay of a single tagged neutral $D$.
\medskip

($2^+$) \underline{$K^- \pi^+(\hp) K^+ \pi^-(-\hp)$.}
\beq
\Gamma^{(+)}(K^-\pi^+, K^+\pi^-)  =
A^4[1 + 2r^2\cos 2\delta + 4r\tilde y - \frac{3}{2}(x^2 - y^2)]~~,
\eeq
where $\tilde y \equiv y \cos \delta +x \sin \delta$.  The correction terms
are probably unmeasurable, so this process serves as a normalization in
comparison with the previous one.
\medskip

($3^+$) \underline{$K^- \pi^+(\hp) S_{\zeta}(-\hp)$.}
\bea
\Gamma^{(+)}(K^-\pi^+, S_\zeta) & = &
A^2 A_{S_\zeta}^2 |1 - \zeta r e^{-i\delta}|^2(1 - 2\zeta y + 3y^2)
\nonumber\\
& \approx & A^2 A_{S_\zeta}^2(1 - 2\zeta r\cos\delta)(1-2\zeta y)~~.
\eea
There thus appears to be ample information to constrain $\cos
\delta$ once $r$ and $y$ are well-enough known.
\medskip

($4^+$) \underline{$K^- \pi^+(\hp)  \ell^- (-\hp)$.}
\beq
\Gamma^{(+)}(K^-\pi^+, \ell^- ) =
A^2 A_{\ell^-}^2 [1 + 2r\tilde y - \frac{3}{2}(x^2 - y^2)]~~.
\eeq
There is a small difference in comparison with the corresponding $C = -1$ case.
\medskip

($5^+$) \underline{$K^- \pi^+(\hp)  \ell^+ (-\hp)$.}
\beq\label{5+}
\Gamma^{(+)}(K^-\pi^+, \ell^+ ) =
A^2 A_{\ell^+}^2 [r^2 + 2ry' + \frac{3}{2}(x^2 + y^2)]~~.
\eeq
In contrast to the corresponding $C=-1$ case, this expression involves
interference between mixing and DCS decay similar to Eq.~(\ref{1+}).
\medskip

($6^+$) \underline{$ S_\zeta(\hp)  \ell^+ (-\hp)$.}
\beq
\Gamma^{(+)}(S_\zeta, \ell^+ )  =
A_{\ell^+}^2 A_{S_\zeta}^2(1 -2\zeta y + 3y^2)~~.
\eeq
This differs from the $C = -1$ case by a term of first order in $y$.
\medskip

An additional set of cases involves the detection of one flavor
eigenstate with direction $\hp$ and a different flavor eigenstate
with direction $-\hp$.  Examples are $K\pi$, $K \rho$, and $K^* \pi$.
Processes with opposite-flavor final states [such as $K^- \pi^+(\hp)
K^+\rho^-(-\hp)$] serve for normalization.  Processes with same-flavor
eigenstates [such as $K^- \pi^+(\hp) K^- \rho^+(-\hp)$]
from $C=-1~D^0 \od$ pairs
then give rise to a leading term in the rate proportional to
$|r_i e^{- i \delta_i} - r_j e^{- i \delta_j}|^2$, where $i$ and $j$
denote two different channels,
$r_i$ is the ratio of DCS to CF amplitudes in the channel $i$, and $\delta_i$
is the strong phase difference between right-sign and wrong-sign decays
in that channel.  Given three such measurements, and independent
determinations of two of the three $r_i$, one can solve for the three phase
differences $\delta_i - \delta_j$ up to discrete ambiguities.  Measurement
of the third $r_i$ reduces those ambiguities by a factor of 2 and provides one
constraint.  There remains an overall ambiguity in the common sign of all
phases.
One can thereby check whether the quantities $\delta_i$ differ from
channel to channel.  The comparison of strengths of $K^* \pi$ and $K \rho$
Dalitz plot bands in (DCS) $D \to K \pi \pi$ and (CF) $D \to \bar K \pi \pi$
decays already indicate the possibility of this difference \cite{ASPC},
since U-spin relations of flavor SU(3) \cite{Uspin} appear to be violated in
such cases.  This method is unable to provide {\it absolute} values of
any of the phases $\delta_i$, in contrast to those based on the examples
$(3^-)$ and $(3^+)$ above.

At this point let us comment briefly on how our results may be modified in the
presence of CP violation. In the Cabibbo-Kobayashi-Maskawa framework CP
violation in neutral $D$ mixing and decays, dominated by the first two
generations, is very small and can be safely neglected. Extensions of the
Standard Model could induce new sources of CP violation. The most likely
sizable effect is a possible new CP violating phase, $\phi = \arg (q\bar
A/pA)$,
occuring in the interference between $D^0-\od$ mixing and decay amplitudes
into $K\pi$ or other hadronic states. Dependence on such a phase requires
mixing and would affect, for instance, the term linear in $t$ in
Eq.~(\ref{eqn:R})
\cite{Berg} and the $ry'$ terms in Eqs.~(\ref{1+}) and (\ref{5+}).
However, it affects Eq.~(\ref{3-}), from which $\cos\delta$ is obtained, only
in terms quadratic in $x,y$, which we neglected.
This is true also for possible CP violation in mixing, $|q/p| \ne 1$.

Other CP violating effects in Eq.~(\ref{3-}) can come from direct CP
violation in Cabibbo-favored $D^0 \to K^-\pi^+$ and in singly
Cabibbo-suppressed $D^0 \to S_\zeta$ decays. This would introduce in the
amplitudes $A$ and $A_{S_\zeta}$ corrections of order $|\mat{K^+\pi^-}{\od}/$
$\mat{K^-\pi^+}
{D^0}| - 1$ and $|\mat{S_\zeta}{\od}/\mat{S_\zeta}{D^0}| -1$, respectively.
Such effects are expected to be very
small in extensions of the Standard Model, where direct CP violation was shown
to be negligible even in DCS decays \cite{BerNir}. This can be checked
directly by comparing $D^0$ and $\od$ branching ratios into CP-conjugate
states. Furthermore, these small effects can be shown to occur only at second
order when combining the rate of Eq.~(\ref{3-}) with its CP-conjugate.

In order to estimate the total sample of events needed to perform a useful
measurement of $\delta$, we note that the rates for the processes of interest
are given in terms of the already measured CF and
DCS rates and the relative strong phase $\delta$.  Let us define an asymmetry
\beq
{\cal A} \equiv \frac{\Gamma^{(-)}(S_+) - \Gamma^{(-)}(S_-)}
{\Gamma^{(-)}(S_+) + \Gamma^{(-)}(S_-)}~~,
\eeq
where $\Gamma^{(-)}(S_\pm)$ is a rate for a $C= -1$ $D^0 \od$ configuration to
decay into a $CP$-eigenstate $S_\pm$ with direction $-\hp$ and a flavor
eigenstate such as $K^- \pi^+$ with direction $+\hp$ [the case $(3^-)$ noted
above)].
Eq.~(\ref{3-}) implies a small asymmetry, ${\cal A} = 2 r \cos \delta$.
For a small asymmetry, a general result is that its error
$\Delta{\cal A}$ is approximately $1/\sqrt{N}$, where $N$ is the total
number of events tagged with CP-even and CP-odd eigenstates. Thus we have
\beq
\Delta (\cos \delta) \simeq \frac{1}{2 r \sqrt{N}}~~.
\eeq
The number $N$ of CP-tagged events decaying to $K^- \pi^+$ is related
to the total number of $D^0 \od$ pairs $N(D^0 \od)$ through $N \approx 0.01
N(D^0 \od){\cal B}(D^0 \to K^-\pi^+) \approx 4 \times 10^{-4} N(D^0 \od)$,
since the branching-ratio-times-efficiency
factor for tagging CP eigenstates is only about 1.1\% \cite{CLEOGW}
(the total branching ratio into CP eigenstates is
larger than about 5\% \cite{PDG}).  With $r = 1.2\tan^2\theta \simeq 0.06$
\cite{CLDCS, FOCD}, one then has
\beq
\Delta(\cos\delta) \approx \frac{400}{\sqrt{N(D^0 \od)}}~~.
\eeq
The cross section for $e^+e^- \to \psi(3770)$ is about 10 nb at the
peak \cite{BES}, while a foreseen integrated luminosity for a charm
factory operating at this energy is about 3 fb$^{-1}$ \cite{CLEOPC}.
One can thus envision collecting $3\times 10^7$ $D\bar D$ pairs, of which half
are charged and half are neutral.  We are entitled to a factor of 2 for
considering both $K^- \pi^+$ and $K^+ \pi^-$ final states.
We thus estimate that one may be able to
reach an accuracy of about 0.07 in $\cos\delta$.

The achievable error on $(x^2 +y^2)^{1/2}$ is of the order of a percent.
To see this, we note that the case $(1^-)$ mentioned above
should yield about $15 \times 10^6 (x^2 +y^2)[{\cal B}(D^0 \to K^- \pi^+)]^2
\simeq 2.2 \times 10^4 (x^2 +y^2)$ $(K^\mp \pi^\pm) (K^\mp \pi^\pm)$
events under the conditions mentioned
above, giving rise to sensitivity to $x^2 +y^2$ at the level of about
$10^{-4}$.  Modest improvements will be possible by adding other final
states.

The parameter $y'= y \cos \delta - x \sin \delta$ is obtained from the $ry'$
terms occuring in the time-dependence (\ref{eqn:R}) in the decay of a single
tagged neutral $D$, and in the rates (\ref{1+}) and (\ref{5+}) measured
by producing $D^0\od$ pairs with $C=+1$. Since $r > x, y$, the accuracy of
measuring $y'$ is expected to be better than of measuring $(x^2 +y^2)^{1/2}$.
Once $\cos\delta$ is measured with the above calculated precision, separate
measurements of $x$ and $y$ at a corresponding level of sensitivity may be
achieved.
\medskip

We thank David Asner, David Cinabro, Jonathan Link, Harry Nelson, Sandip
Pakvasa,
and Alex Smith for useful discussions, and A. I. Sanda for extending the
hospitality of Nagoya University during part of this work.
This work was supported in part by the United States Department of
Energy through Grant No.\ DE FG02 90ER40560, by the Israel Science Foundation
founded by the Israel Academy of Sciences and Humanities,
and by the U. S. -- Israel Binational Science Foundation through Grant
No.\ 98-00237.

\def \ajp#1#2#3{Am.\ J. Phys.\ {\bf#1} (#3) #2}
\def \apny#1#2#3{Ann.\ Phys.\ (N.Y.) {\bf#1} (#3) #2}
\def \app#1#2#3{Acta Phys.\ Polonica {\bf#1} (#3) #2}
\def \arnps#1#2#3{Ann.\ Rev.\ Nucl.\ Part.\ Sci.\ {\bf#1} (#3) #2}
\def \art{and references therein}
\def \cmts#1#2#3{Comments on Nucl.\ Part.\ Phys.\ {\bf#1} (#3) #2}
\def \cn{Collaboration}
\def \cp89{{\it CP Violation,} edited by C. Jarlskog (World Scientific,
Singapore, 1989)}
\def \efi{Enrico Fermi Institute Report No.\ }
\def \epjc#1#2#3{Eur.\ Phys.\ J. C {\bf#1} (#3) #2}
\def \f79{{\it Proceedings of the 1979 International Symposium on Lepton and
Photon Interactions at High Energies,} Fermilab, August 23-29, 1979, ed. by
T. B. W. Kirk and H. D. I. Abarbanel (Fermi National Accelerator Laboratory,
Batavia, IL, 1979}
\def \hb87{{\it Proceeding of the 1987 International Symposium on Lepton and
Photon Interactions at High Energies,} Hamburg, 1987, ed. by W. Bartel
and R. R\"uckl (Nucl.\ Phys.\ B, Proc.\ Suppl., vol.\ 3) (North-Holland,
Amsterdam, 1988)}
\def \ib{{\it ibid.}~}
\def \ibj#1#2#3{~{\bf#1} (#3) #2}
\def \ichep72{{\it Proceedings of the XVI International Conference on High
Energy Physics}, Chicago and Batavia, Illinois, Sept. 6 -- 13, 1972,
edited by J. D. Jackson, A. Roberts, and R. Donaldson (Fermilab, Batavia,
IL, 1972)}
\def \ijmpa#1#2#3{Int.\ J.\ Mod.\ Phys.\ A {\bf#1} (#3) #2}
\def \ite{{\it et al.}}
\def \jhep#1#2#3{JHEP {\bf#1} (#3) #2}
\def \jpb#1#2#3{J.\ Phys.\ B {\bf#1} (#3) #2}
\def \lg{{\it Proceedings of the XIXth International Symposium on
Lepton and Photon Interactions,} Stanford, California, August 9--14 1999,
edited by J. Jaros and M. Peskin (World Scientific, Singapore, 2000)}
\def \lkl87{{\it Selected Topics in Electroweak Interactions} (Proceedings of
the Second Lake Louise Institute on New Frontiers in Particle Physics, 15 --
21 February, 1987), edited by J. M. Cameron \ite~(World Scientific, Singapore,
1987)}
\def \kdvs#1#2#3{{Kong.\ Danske Vid.\ Selsk., Matt-fys.\ Medd.} {\bf #1},
No.\ #2 (#3)}
\def \ky85{{\it Proceedings of the International Symposium on Lepton and
Photon Interactions at High Energy,} Kyoto, Aug.~19-24, 1985, edited by M.
Konuma and K. Takahashi (Kyoto Univ., Kyoto, 1985)}
\def \mpla#1#2#3{Mod.\ Phys.\ Lett.\ A {\bf#1} (#3) #2}
\def \nat#1#2#3{Nature {\bf#1} (#3) #2}
\def \nc#1#2#3{Nuovo Cim.\ {\bf#1} (#3) #2}
\def \nima#1#2#3{Nucl.\ Instr.\ Meth. A {\bf#1} (#3) #2}
\def \np#1#2#3{Nucl.\ Phys.\ {\bf#1} (#3) #2}
\def \os{XXX International Conference on High Energy Physics, 27 July
-- 2 August 2000, Osaka, Japan}
\def \PDG{Particle Data Group, D. E. Groom \ite, \epjc{15}{1}{2000}}
\def \pisma#1#2#3#4{Pis'ma Zh.\ Eksp.\ Teor.\ Fiz.\ {\bf#1} (#3) #2 [JETP
Lett.\ {\bf#1} (#3) #4]}
\def \pl#1#2#3{Phys.\ Lett.\ {\bf#1} (#3) #2}
\def \pla#1#2#3{Phys.\ Lett.\ A {\bf#1} (#3) #2}
\def \plb#1#2#3{Phys.\ Lett.\ B {\bf#1} (#3) #2}
\def \pr#1#2#3{Phys.\ Rev.\ {\bf#1} (#3) #2}
\def \prc#1#2#3{Phys.\ Rev.\ C {\bf#1} (#3) #2}
\def \prd#1#2#3{Phys.\ Rev.\ D {\bf#1} (#3) #2}
\def \prl#1#2#3{Phys.\ Rev.\ Lett.\ {\bf#1} (#3) #2}
\def \prp#1#2#3{Phys.\ Rep.\ {\bf#1} (#3) #2}
\def \ptp#1#2#3{Prog.\ Theor.\ Phys.\ {\bf#1} (#3) #2}
\def \rmp#1#2#3{Rev.\ Mod.\ Phys.\ {\bf#1} (#3) #2}
\def \rp#1{~~~~~\ldots\ldots{\rm rp~}{#1}~~~~~}
\def \si90{25th International Conference on High Energy Physics, Singapore,
Aug. 2-8, 1990}
\def \slc87{{\it Proceedings of the Salt Lake City Meeting} (Division of
Particles and Fields, American Physical Society, Salt Lake City, Utah, 1987),
ed. by C. DeTar and J. S. Ball (World Scientific, Singapore, 1987)}
\def \slac89{{\it Proceedings of the XIVth International Symposium on
Lepton and Photon Interactions,} Stanford, California, 1989, edited by M.
Riordan (World Scientific, Singapore, 1990)}
\def \smass82{{\it Proceedings of the 1982 DPF Summer Study on Elementary
Particle Physics and Future Facilities}, Snowmass, Colorado, edited by R.
Donaldson, R. Gustafson, and F. Paige (World Scientific, Singapore, 1982)}
\def \smass90{{\it Research Directions for the Decade} (Proceedings of the
1990 Summer Study on High Energy Physics, June 25--July 13, Snowmass,
Colorado),
edited by E. L. Berger (World Scientific, Singapore, 1992)}
\def \tasi{{\it Testing the Standard Model} (Proceedings of the 1990
Theoretical Advanced Study Institute in Elementary Particle Physics, Boulder,
Colorado, 3--27 June, 1990), edited by M. Cveti\v{c} and P. Langacker
(World Scientific, Singapore, 1991)}
\def \yaf#1#2#3#4{Yad.\ Fiz.\ {\bf#1} (#3) #2 [Sov.\ J.\ Nucl.\ Phys.\
{\bf #1} (#3) #4]}
\def \zhetf#1#2#3#4#5#6{Zh.\ Eksp.\ Teor.\ Fiz.\ {\bf #1} (#3) #2 [Sov.\
Phys.\ - JETP {\bf #4} (#6) #5]}
\def \zpc#1#2#3{Zeit.\ Phys.\ C {\bf#1} (#3) #2}
\def \zpd#1#2#3{Zeit.\ Phys.\ D {\bf#1} (#3) #2}

\end{document}